# Algorithm-Based Linearly Graded Compositions of GeSn on GaAs (001) via Molecular Beam Epitaxy

Calbi Gunder[1*], Mohammad Zamani-Alavijeh[2], Emmanuel Wangila[1], Fernando Maia de Oliveira[3], Aida Sheibani[2], Serhii Kryvyi[3], Paul C. Attwood[4], Yuriy I. Mazur[3], Shui-Qing Yu[3,5], Gregory J. Salamo[2,3]

[1]*Materials Science and Engineering, University of Arkansas, Fayetteville, AR 72701, USA*
[2]*Department of Physics, University of Arkansas, Fayetteville, AR 72701, USA*
[3]*Institute for Nanoscience and Engineering, University of Arkansas, Fayetteville, AR 72701, USA*
[4]*Gunder and Attwood Armories, El Dorado Springs, MO 64744, USA*
[5]*Department of Electrical Engineering, University of Arkansas, Fayetteville, AR 72701, USA*

**Abstract**

The growth of high-composition GeSn films of the future will likely be guided via algorithms. In this study we show how a logarithmic-based algorithm can be used to obtain high-quality GeSn compositions up to 16 % on GaAs (001) substrates via molecular beam epitaxy. Within we demonstrate composition targeting and logarithmic gradients to achieve linearly graded pseudomorph $Ge_{1-x}Sn_x$ compositions up to 10 % before partial relaxation of the structure and a continued gradient up to 16 % GeSn. In this report, we use X-ray diffraction, simulation, SIMS and atomic force microscopy to analyze and demonstrate some of the possible growths that can be produced with the enclosed algorithm. This methodology of growth is a major step forward in the field of GeSn development and the first demonstration of algorithmically driven, linearly graded GeSn films.

**Keywords:** GeSn; algorithm-based GeSn growth; logarithmic-based GeSn growth

*Corresponding Author, E-mail: calbigunder@gmail.com

## Introduction

The controlled and reproducible growth of germanium–tin (GeSn) films through molecular beam epitaxy (MBE) holds paramount significance for the advancement of GeSn research. The unique combination of a tunable bandgap, high carrier mobility, and compatibility with silicon technology makes GeSn systems an attractive choice for optoelectronic devices such as high-efficiency photodetectors, lasers, and infrared sensors [1–3]. When alloying GeSn into a direct bandgap material, there are many factors that need to be taken into consideration. This includes the strain state of the material, the amount of Sn incorporation, and the short-range ordering within the GeSn lattice structure [4–6]. All three factors become important when determining the required specifications of a GeSn film for device applications in the future.

In an era where the quest for efficient materials underpins the evolution of artificial intelligence (AI) platforms, achieving precision in material growth becomes not just a scientific endeavor but also a cornerstone of technological innovation for the next generation of semiconductor devices [7]. In particular, the pursuit of controlled GeSn growth guided by algorithmic principles offers a tantalizing prospect. By taking advantage of the power of MBE's monolayer control, it is possible to meticulously tailor the composition of GeSn films while minimizing defect density and maximizing Sn composition. The control of material features, such as the in-plane/out-of-plane strain, composition, and even the possibility of controlling short-range ordering, would result in unparalleled control over the resultant bandgap of GeSn and represent a promising field of study with implications that resonate deeply in AI technology. Indeed, algorithm-based GeSn growth represents more than just a scientific curiosity, as it opens the door to a profound shift in material engineering.

In this paper, we introduce a logarithmic-based function that is used to linearize the change in flux or the equivalent beam equivalent pressure (BEP) and thus linearize the change in Sn composition across $Ge_{1-x}Sn_x$ gradients. By demonstrating the feasibility of cultivating a linear gradient of $Ge_{1-x}Sn_x$ structures, this study opens avenues for novel and transformative projects capable of generating algorithms that can manipulate the curvature of the alloying profile. Through the application of the enclosed logarithmic-based function, the potential arises to craft optimized GeSn structures. These equations can be integrated with GeSn alloying simulations to explore the most effective method for distributing Sn across specific thicknesses, aiming to achieve a direct bandgap material tailored to a specific wavelength of interest. The versatility of the logarithmic-based function extends beyond gradient growth, as it can also be employed for selecting a particular composition or stepped composition structures. It also allows for a future study to identify and utilize the role of strain gradient and strain energy to achieve a higher Sn content without dislocations. For example, the strain gradient can determine the allowed Sn content per unit length during growth, while the accumulated strain energy can determine the limiting Sn content before relaxation by dislocations. In this report, however, we show how the calibration of the growth function can occur and demonstrate the function through growths targeting specific compositions of GeSn. We also present a new analytical tool for easily calculating $Ge_{1-x}Sn_x$ compositions using reciprocal space map (RSM) data, which is very useful in the field of GeSn epitaxial growths.

**Experimental details**

The GeSn films were grown using an MBE Riber-32 system. This study used undoped GaAs (001) substrates purchased from Wafer Technology LTD, located in Milton Keynes, UK. The GaAs wafers were degassed at 300 °C for 2 h before being transferred through a vacuum transfer line into the main chamber for oxide removal under arsenic flux. Each GaAs (001) wafer received a 230 nm buffer grown at 585 °C, which was monitored via a bandit system. For these growths, we maintained an arsenic-to-gallium ratio of 15:1. For all GeSn samples grown, a constant Ge BEP of $2.376 \times 10^{-7}$ Torr was used, along with a manipulator temperature gradient ranging from 200 to 50 °C at a ramp rate of 10 °C/min, using the same procedure reported previously by Gunder et al. [8]. In addition to this, an extra 7 min of growth was added to the conclusion of each logarithmic-based growth at the final Sn cell temperature ($T_{Sn}$). This addition aims to illustrate the growth behavior of the final layer of GeSn after the gradient.

For analysis, we relied on atomic force microscopy (AFM), X-ray diffraction (XRD), and secondary ion mass spectrometry (SIMS) as the primary characterization tools. The AFM system was a D3100 Nanoscope V made by Bruker and manufactured in the USA (Billerica, MA), in which high-quality tips (HQ.NSC15/Al BS made by MikroMasch located in Watsonville, CA of the USA) were used for tapping mode measurements to observe the surface roughness and morphology. XRD was used to observe the strain profile and GeSn compositions via reciprocal space mapping (RSM) and omega-2theta scans of the grown films. The machine used a Panalytical X'Pert MRD diffractometer from The Netherlands (Almelo) utilizing a CuKα1 source (λ = 0.15406 nm) with a four-bounce Ge (220) monochromator and a multilayer focusing mirror. For the high-intensity configuration, a pixel detector was used. A three-bounce Ge (220) channel cut analyzer in front of a standard proportional detector was used for the high-resolution configuration. SIMS characterization was performed at EAG Laboratories to analyze the alloying profile of the three films grown.

In order to grow linearly graded GeSn films, it is necessary to provide a linearly increasing flux or equivalent BEP of Sn during the growth. This requires an understanding of how the BEP of Sn ($P_{Sn}$) changes with respect to a change in Sn cell temperature ($T_{Sn}$). This can be described as an exponential relationship with reference to a change in temperature, as shown in Equation (1) [9]:

$$P_{Sn} = P_0 \, e^{\frac{-\Delta H_e}{RT_{Sn}}} \qquad (Eq.\,1)$$

where R is the gas constant, ($\Delta H_e$) is the molar heat of evaporation, and ($P_0$) represents the constant of integration. Both ($\Delta H_e$) and ($P_0$) depend on material properties, which are ascertained through fitting Equation (1) to a ($P_{Sn}$) versus ($T_{Sn}$) plot taken for the Sn cell across a range of temperatures using a flux gauge. An example of this plot can be observed in Figure 1.

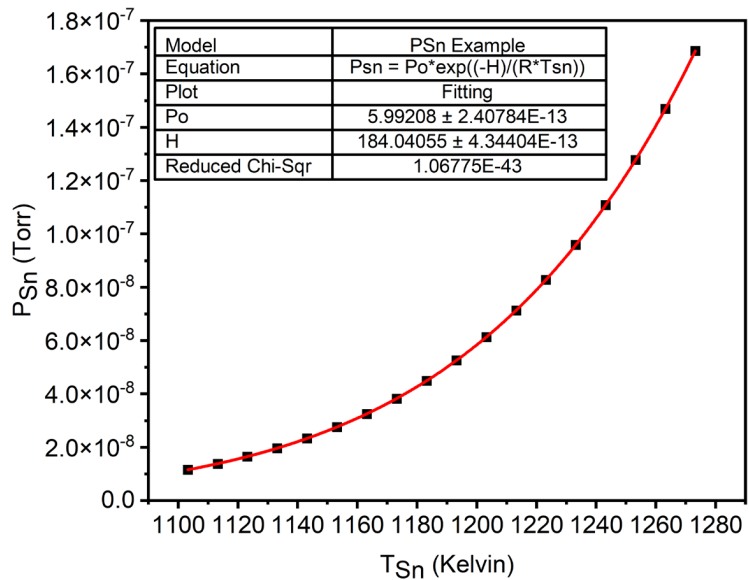

Fig. 1: Example of BEP of Sn versus Sn cell temperature plot for finding coefficients $\Delta H_e$ and $P_0$.

The logarithmic function used in this paper is given by Equation (2), where the BEP of Sn is designed to follow a linear equation with respect to growth time ($t$), as expressed by $at + b$.

The variable $(a)$ represents the rate of BEP change across time, while $(b)$ is the offset variable that sets the starting BEP at $t = 0$. Alternatively, $(b)$ can be equated to $(P_{Sn})$ and solved directly through Equation (1) by setting $(T_{Sn})$ to the desired starting temperature.

$$T_{Sn} = \frac{-\Delta H_e}{R \ln\left(\frac{at + b}{P_0}\right)} \qquad (Eq.\,2)$$

When applying this algorithm, we consider $(T_{Sn})$ as the base Sn cell temperature of our dual-filament effusion cell, while the tip temperature is set separately to follow its change by a +50 K offset. The values of the variables used in Equation (2) for the growth of these samples are shown in Table 1. The user defined $(a)$ rate change of Sn BEP was selected as $5.1943 \times 10^{-10}\ \frac{Torr}{min}$ to allow for a 15.33 K effusion cell ramp rate change for the first minute of the logarithmic function. This was initially selected to allow room for a future growth study of different grading rates, such as a $2a$ rate and a $0.5a$ rate.

Composition targeting can be achieved through growth calibration and the study of the corresponding XRD data. For example, the beginning of our gradient is $T_{Sn} = 1093.15$ K, which is near $Ge_{0.998}Sn_{0.002}$, and the end of the gradients corresponds to $T_{Sn} = 1273.15$ K and $T_{Sn} = 1298.15$ K, with compositions near $Ge_{0.9}Sn_{0.1}$ and $Ge_{0.84}Sn_{0.16}$, respectively, as shown in Table 1. These composition data points can be used to replace $P_{Sn}$ in the pressure versus temperature plot and fitted again to predict composition versus $T_{Sn}$. SIMS data may also be used to increase the accuracy of composition targeting; however, it is still advised to couple it with XRD in order to distinguish crystal structure with reference to composition.

| Table 1: Coefficients/variables and $T_{Sn}$ temperature range used in growths. | | |
|---|---|---|
| Variables/Coefficients | Values | |
| R | $8.314459 \times 10^{-3}\ \frac{kJ}{mol * K}$ | |
| a | $5.1943 \times 10^{-9}\ \frac{Torr}{min}$ | |
| b | $9.62999 \times 10^{-9}\ Torr$ | |
| $\Delta H_e$ | $184.04055\ \frac{kJ}{mol}$ | |
| $P_0$ | $5.99208\ Torr$ | |
| Samples | $T_{Sn}$ at start (K) | $T_{Sn}$ at end (K) |
| 4% | 1093.15 | 1207.85 |
| 10% | | 1273.15 |
| 16% | | 1298.15 |

**Results and discussion**

Three samples were grown using the logarithmic-based function to linearize the change in Sn flux, resulting in samples with final crystalline compositions near 4%, 10%, and 16%. The Sn composition of each sample was determined from RSM data by defining the evidenced lattice mismatch as a function of the local Sn content at each lattice location [10]. To do so, the relaxed lattice parameter $a_0^{GeSn}$ of the GeSn alloy and its set of elastic constants, $C_{11}^{GeSn}$ and $C_{12}^{GeSn}$, can be described in terms of Vegard's law as follows:

$$a_0^{GeSn} = xa_0^{Sn} + (1-x)a_0^{Ge} \qquad Eq.(3a)$$

$$C_{11,12}^{GeSn} = xC_{11,12}^{Sn} + (1-x)C_{11,12}^{Ge} \qquad Eq.(3b)$$

The value of each constant is presented in Table 2.

Table 2: Relaxed lattice parameters and elastic constants of Sn and Ge, as well as the functions $\alpha$ and $\beta$ used in Eq. (4)

| | |
|---|---|
| $a_0^{Sn}(nm)$ | 0.6489 |
| $a_0^{Ge}(nm)$ | 0.5658 |
| $C_{11}^{Sn}(GPa)$ | 69 |
| $C_{12}^{Sn}(GPa)$ | 29.3 |
| $C_{11}^{Ge}(GPa)$ | 126 |
| $C_{12}^{Ge}(GPa)$ | 44 |
| $\alpha$ | $2.0474 a_\parallel^{GeSn} + 3.96945 a_\perp^{GeSn} - 2.16591$ |
| $\beta$ | $12.25654 a_\parallel^{GeSn} + 17.54914 a_\perp^{GeSn} - 16.86405$ |

By solving Eq. 3 for $x$, we are presenting a new equation to calculate Sn content from the components of the scattering vectors $Q_x$ and $Q_z$ by:

$$x = \alpha - \sqrt{\alpha^2 - \beta} \qquad Eq.(4)$$

where the parameters $\alpha$ and $\beta$ are functions of the in-plane and out-of-plane lattice parameters $a_\parallel^{GeSn}$ and $a_\perp^{GeSn}$, as shown in Table 2, which are given by $a_\parallel^{GeSn} = 2\pi\sqrt{(h^2+k^2)/Q_x^2}$ and $a_\perp^{GeSn} = 2\pi l/Q_z$ for an asymmetrical $hkl$ reflection. A detailed description of how to derive Equation (4) is shown in the supporting Supplementary Information.

The RSM shown in Figure 2d–f indicates that the in-plane lattice parameters $a_\parallel^{GeSn}$ remain nearly unchanged for compositions of Sn up to 8–10%, indicating a pseudomorphic growth of GeSn [11]. The pseudomorphic region is likely the result of managing the Sn distribution across the thickness and its resultant effect on the strain, leading to stabilization of the parallel lattice parameters $a_\parallel^{GeSn}$. This phenomenon is likely attributed to the interplay between composition and its associated critical thickness, suggesting a correlation that enables the system to sustain such strain characteristics even amid substantial changes in composition. The addition of 7 min of GeSn growth at the end shows that the 4% sample continues to grow fully strained. The next sample, which continues the function to 10%, begins to show the start of a relaxation process, as seen in the RSM depicted in Figure 2e. The same pseudomorphic growth is observed for much of the film until the composition approaches 10%, which is evidenced by the expansion of the $a_\parallel^{GeSn}$ lattice parameters. At this point, the 7 min growth at the end becomes more apparent, and it is observed that the structure begins to relax, which can be attributed to the curvature in the tail of the RSM. The last sample grown with a crystal composition near 16% shows that the structure stops its expansion of the out-of-plane lattice parameter $a_\perp^{GeSn}$, while its in-plane lattice parameters $a_\parallel^{GeSn}$ begin to expand in order to continue adding Sn to its structure. This is evidenced in Figure 2f, where the elongation line is expanding consistently parallel to the $Q_x$ axis, while $Q_z$ maintains its position. This suggests that two types of GeSn regions appear for the 16% sample: fully strained (pseudomorphic) and a graded relaxation region up to 60%.

Simulations consisting of (004) omega-2theta scans were carried out to define the Sn depth distribution and to confirm the maximum Sn concentration. The simulations were performed in Epitaxy 4.2 software from Panalytical, taking into account the strain state of the layers according to XRD-measured RSM. The best fit for each sample was obtained for a linear Sn depth profiling distribution in the GeSn layers. The initial and final Sn concentrations are presented in Table 3. As can be seen, thickness fringes are observed in the measured XRD omega-2theta scan for the 4% sample, allowing us to use Equation (5), where λ is the X-ray wavelength, θ is the peak position, and n is the peak order. This allows the calculation of the estimated $Ge_{1-x}Sn_x$ total layer thickness, resulting in $t_{Ge_{1-x}Sn_x} = 112 \pm 5$ nm [12].

$$t_{Ge_{1-x}Sn_x} = \frac{(n_1 - n_2)\lambda}{2(\sin\theta_1 - \sin\theta_2)} \qquad Eq.\,(5)$$

The simulated results of this growth also agree well with this thickness calculation and allow the separation of the approximate thicknesses for the 18 min $Ge_{1-x}Sn_x$ gradient and the additional 7 min worth of growth at composition, resulting in a thickness of $75 \pm 5$ and $35 \pm 5$ nm, respectively, for a sum thickness near 110 nm. The simulation also agrees with the maximum composition values calculated from Equation (4) for the three samples. In the case of the 16% sample, the simulation was performed according to RSM by creating a model consisting of a fully strained layer followed by a partly relaxed GeSn layer. The best fit was obtained for a maximum of $10 \pm 0.5\%$ Sn concentration in the pseudomorphic region and about $15.5 \pm 0.5\%$ for the partly relaxed region.

| Table 3: Simulation results of GeSn films. | | | |
|---|---|---|---|
| Sample | Layer | Initial Sn concentration (%) | Final Sn concentration (%) |
| 4% | Pseudomorphic | 0 ± 0.5 | 3.9 ± 0.5 |
| 10% | Pseudomorphic | 0.5 ± 0.5 | 9.8 ± 0.5 |
| 16% | Pseudomorphic | 0.5 ± 0.5 | 10 ± 0.5 |
| | Partly relaxed up R ≈ 60% | - | 15.5 ± 0.5 |

Another feature that can be observed from the RSM of the three samples is that each subsequent sample with a higher Sn composition also results in a larger $Q_x$ width for the pseudomorphic region. This could be due to defect propagation or related to the act of growing additional material at the higher compositions along the gradient (larger lattice constants), causing additional strain to be applied to the underlining lattice structure, resulting in the in-plane lattice constants of the lower compositions expanding. A similar trend in which the FWHM of rocking curves increases within the pseudomorphic region with respect to the three samples was observed. The film quality will be discussed later by examining the corresponding rocking curves.

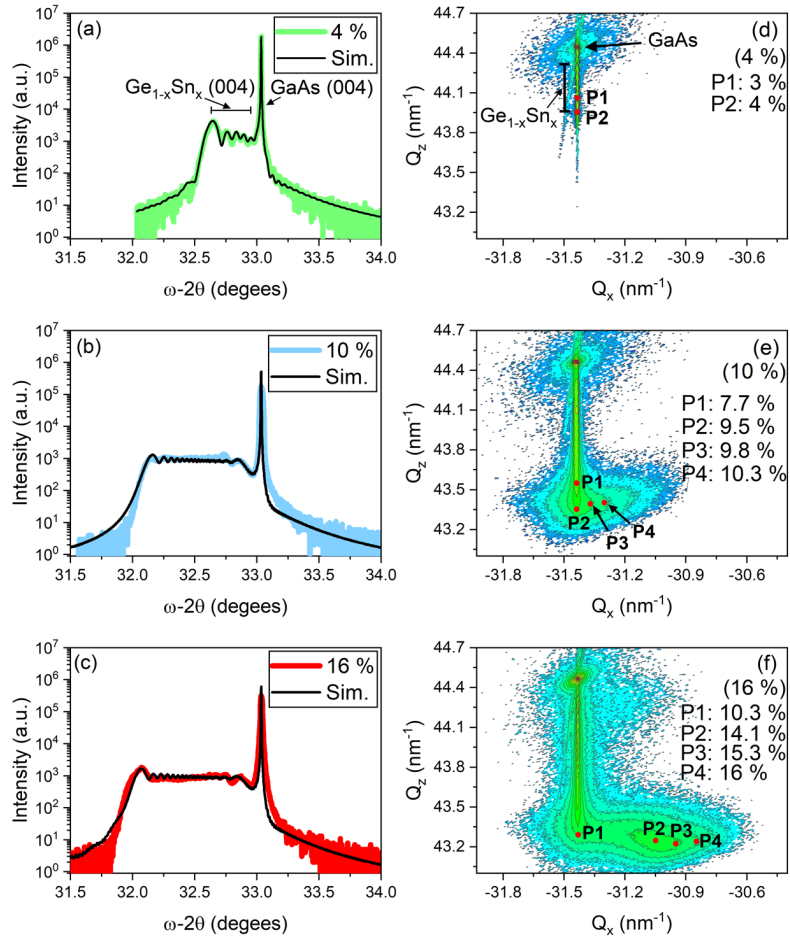

Fig. 2: XRD and simulation of the 4, 10, and 16% samples. Plots (**a**–**c**) consist of both omega-2theta (004) scans and simulation, while (**d**–**f**) consist of the RSM from the $\bar{2}\bar{2}4$ crystal planes. The XRD RSM intensity change is observed as a gradient in color from the highest intensity (red) to the lowest intensity (blue). All RSM Sn percentage positions labeled as red points were calculated with Equation (4).

SIMS characterization was conducted to confirm the alloying of these gradients, as shown in Figure 3. From the SIMS etching profile, it is clear that the $Ge_{1-x}Sn_x$ composition is changing linearly for the majority of the growth using our logarithmic algorithm to control the Sn cell temperature. Starting with the 4% sample, it can be observed that the linear gradient grows to 3%, as seen in Figure 3a, and that within the last 7 min of growth, an increase in the $Ge_{1-x}Sn_x$ crystal structure composition near the 4% mark is observed from the XRD results, as presented in Figure 2a,d. The linear grading profile is also observed in samples 10% and 16% in Figure 3b,c. In each of the samples, an increase in Sn content exceeding that of the GeSn composition calculations from XRD occurs. It is likely that this excess Sn was not incorporated into the GeSn crystal structure, as evidenced by the surface morphology, which will be discussed later within the paper.

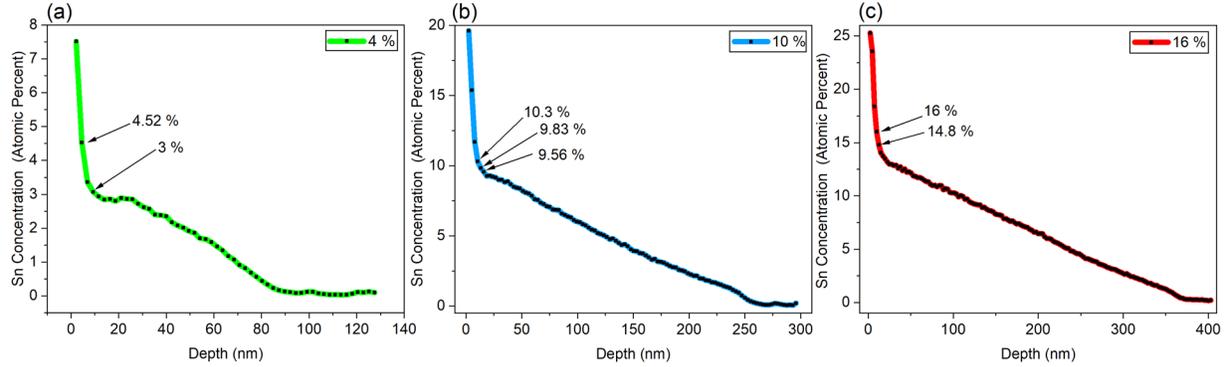

Fig. 3: SIMS Sn concentration versus depth. (a) Sample 4%, (b) 10% and (c) 16%.

The in-plane $\varepsilon_{\parallel}^{GeSn}$, out-of-plane $\varepsilon_{\perp}^{GeSn}$ strain values for the three samples were calculated using Equation (6a,b), where $a_{\parallel}^{GeSn}$ and $a_{\perp}^{GeSn}$ are coupled with the ideal relaxed lattice constant $a_0^{GeSn}$ of its corresponding GeSn composition.

$$\varepsilon_{\parallel}^{GeSn} = \frac{a_{\parallel}^{GeSn} - a_0^{GeSn}}{a_0^{GeSn}} \qquad Eq.\,(6a)$$

$$\varepsilon_{\perp}^{GeSn} = \frac{a_{\perp}^{GeSn} - a_0^{GeSn}}{a_0^{GeSn}} \qquad Eq.\,(6b)$$

Using Equations (4) and (6a,b), a 3D plot of $\varepsilon_{\parallel}^{GeSn}$ and $\varepsilon_{\perp}^{GeSn}$ strain versus composition can be created to more easily observe the change in strain versus composition, as shown in Figure 4. Within this plot, red points are inset to match the corresponding RSM composition points in Figure 2. The corresponding strain and compositional data are presented in Table 4. From these data sets, it can be observed that the in-plane and out-of-plane strain states appear to maximize near −0.015

(−1.5%) and 0.01 (1%), respectively. This is observed in both samples: 10% at position P2 and 16% at position P1.

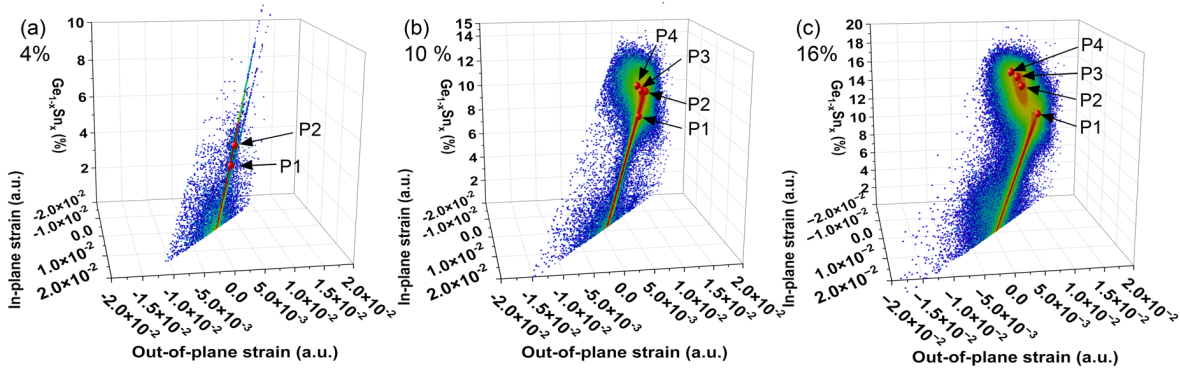

Fig. 4: Plots of (**a**) 4%, (**b**) 10%, and (**c**) 16% samples representing XRD RSM translated into in-plane and out-of-plane strains versus composition. The inset red points correspond to the composition points from the RSM in Figure 2. These plots were generated using Equations (4) and (6a,b).

| Table 4: Composition and its corresponding in-plane and out-of-plane strains. ||||| 
|---|---|---|---|---|
| Sample | Position | Composition (%) from Eq. 4 | In-plane strain (a. u.) from Eq. 6a | Out-of-plane strain (a. u.) from Eq. 6b |
| 4 % | P1 | 3 | -0.0052 | 0.0037 |
|  | P2 | 4 | -0.0066 | 0.0046 |
| 10 % | P1 | 7.7 | -0.0122 | 0.0086 |
|  | P2 | 9.5 | -0.0148 | 0.0104 |
|  | P3 | 9.8 | -0.013 | 0.0092 |
|  | P4 | 10.3 | -0.0114 | 0.0081 |
| 16 % | P1 | 10.3 | -0.01552 | 0.011 |
|  | P2 | 14.1 | -0.009 | 0.0064 |
|  | P3 | 15.3 | -0.0074 | 0.0053 |
|  | P4 | 16 | -0.0051 | 0.0036 |

The film quality was evaluated using (004) omega rocking curves in the high-resolution configuration. The omega scans were collected in increments of 0.5 degrees across the whole omega-2theta range of the gradient, as depicted in Figure 5. The crystal quality of the epitaxial layers was estimated by obtaining the full width at half maximum (FWHM) for omega rocking curve scans. Near the omega-2theta 32.5-degree mark, it can be seen that the rocking curves begin to form two peaks. An example of this is shown and fitted in Figure 5b, in which the narrow peak is related to the pseudomorphic region while the broader peak originates from the partially relaxed region. Results from these fittings suggest that for all samples, the pseudomorphic region maintains extremely high-quality levels, ranging from 30–50 arcseconds in the entire omega-2theta range. This compares to about 16 arcseconds for the GaAs substrate. At the same time, the trend of increasing FWHM follows the relaxation process toward the decreasing omega-2theta angle (increasing Sn content). For the 16% sample, it can be seen that an increase in the FWHM in the relaxation region occurs compared to the pseudomorphic region. This decrease in quality is likely due to the relaxation process and defect propagation in the film. These results directly support the claim that gradient-based structures can help mitigate or suppress defect propagation [8].

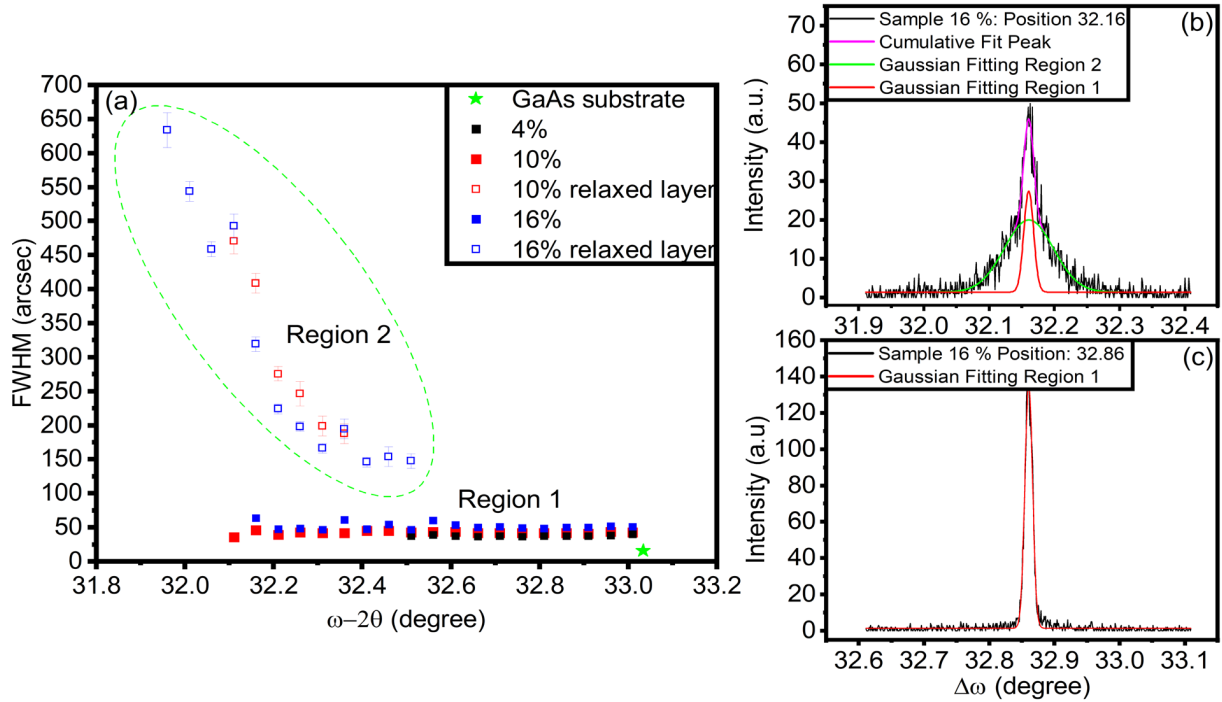

Fig. 5: FWHM omega rocking curves of the GeSn gradients in arcseconds. Plot (**a**) contains all of the FWHM values from gaussian fittings of both relaxing region 2 and pseudomorphic region 1. Plots (**b**,**c**) show how gaussian fittings were used to fit each region.

The surface characteristics of the GeSn films depict similar surface features, with one notable exception. In the 16% sample, some observable Sn segregation sites begin to sparsely appear across the surface of the sample, with normal regions in between, as shown in Figure 6. Within the normal regions small dots are present on the surface which could be a sign of excess

Sn. The 4%, 10%, and 16% samples have an average surface roughness of 0.337, 1.78, and 2.8 nm, respectively.

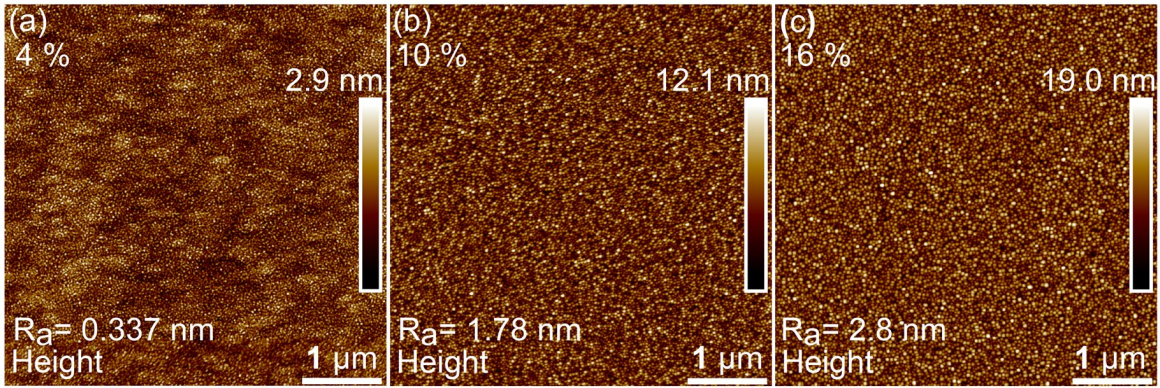

Fig. 6: AFM 5 × 5 µm² images representing the surface morphology of the (**a**) 4%, (**b**) 10%, and (**c**) 16% samples.

A likely solution for improving the surface quality even further is to set up a system capable of continuously monitoring the surface temperature of the film with respect to the manipulator thermal couple temperature during growth while having the ability to not only heat the manipulator but also to directly cool it. By coupling this type of system with a eutectic composition tracking algorithm designed to maintain an optimum relationship between the surface temperature of the GeSn film and the specific Sn composition that is being grown at a specific point in time while also relating the algorithm to the sticking coefficient to aid in controlling the incorporation rate of Sn into GeSn, the crystal quality may be significantly improved while achieving higher composition films. Alternatively, it may be possible to use a dummy effusion cell to radiatively heat the surface of the sample to create a temperature gradient between the top of the film and the substrate to aid in crystallizing the GeSn film even further.

**Conclusion**

In this study, we demonstrated a new log-based algorithm for controlled MBE growth of high-quality, linearly graded $Ge_{1-x}Sn_x$ films. GeSn films with crystal qualities near those of the substrate were achieved. This report also presents an additional XRD tool that allows for the calculation of GeSn compositions through RSM data. This growth algorithm is applicable to the growth of GeSn on many different substrate systems, such as germanium, silicon, sapphire, indium arsenide, InGaAs, and silicon carbide, just to name a few. This methodology can also help to determine the approximate BEP ratio of Sn to Ge in order to obtain specific compositions through a calibration process.